\documentclass[preprint,5p,times,authoryear]{elsarticle}
\journal{Physics of the Dark Universe}

\usepackage[utf8]{inputenc} % Required for inputting international characters
\usepackage[T1]{fontenc}
\usepackage{ae,aecompl}
\usepackage{graphicx}	% Including figure files
\usepackage{amsmath}	% Advanced maths commands
\usepackage{amssymb}	% Extra maths symbols
\usepackage{cuted}
\usepackage{hyperref}
\usepackage{cleveref}
\usepackage{color}
\usepackage{ulem}

\usepackage[all]{nowidow}
\makeatletter
\providecommand{\doi}[1]{%
  \begingroup
    \let\bibinfo\@secondoftwo
    \urlstyle{rm}%
    \href{http://dx.doi.org/#1}{%
      doi:\discretionary{}{}{}%
      \nolinkurl{#1}%
    }%
  \endgroup
}
\makeatother
\hypersetup{
	colorlinks,
	linkcolor={red!50!black},
	citecolor={magenta},
	urlcolor={blue!80!black}
}
\def \MBH 			{M_{\rm BH}}

\def \Mtot 			{M_{\rm tot}}

\begin{document}
\input{aas_macros.sty}

\begin{frontmatter}
\title{Novel constraints on fermionic dark matter from galactic observables I:\\The Milky Way}
	
\author[cct,icranet]{C.~R.~Argüelles}
\ead{carlos.arguelles@icranet.org}
\author[icranet,icra,nice]{A.~Krut}
\ead{andreas.krut@icranet.org}
\author[icranet,icra,cbpf]{J.~A.~Rueda}
\ead{jorge.rueda@icra.it}
\author[icranet,icra,cbpf]{R.~Ruffini}
\ead{ruffini@icra.it}

\address[cct]{Instituto de Astrofísica de La Plata (CCT La Plata, CONICET, UNLP), Paseo del Bosque, B1900FWA La Plata, Argentina}
\address[icranet]{ICRANet, Piazza della Repubblica 10, I--65122 Pescara, Italy}
\address[icra]{Dipartimento di Fisica and ICRA, Sapienza Università di Roma, P.le Aldo Moro 5, I--00185 Rome, Italy}
\address[nice]{University of Nice-Sophia Antipolis, 28 Av. de Valrose, 06103 Nice Cedex 2, France}
\address[cbpf]{ICRANet-Rio, CBPF, Rua Dr.~Xavier Sigaud 150, Rio de Janeiro, RJ, 22290--180, Brazil}

\begin{abstract}
We have recently introduced a new model for the distribution of dark matter (DM) in galaxies based on a self-gravitating system of massive fermions at finite temperatures, the Ruffini-Argüelles-Rueda (RAR) model. We show that this model, for fermion masses in the keV range, explains the DM halo of the Galaxy and predicts the existence of a denser quantum core at the center. We demonstrate here that the introduction of a cutoff in the fermion phase-space distribution, necessary to account for the finite Galaxy size, defines a new solution with a central core which represents an alternative to the black hole (BH) scenario for SgrA*. For a fermion mass in the range $mc^2 = 48$ -- $345$~keV, the DM halo distribution is in agreement with the Milky Way rotation curve data, while harbors a dense quantum core of about $4\times10^6 M_\odot$ within the S2-star pericenter. 
\end{abstract}


\end{frontmatter}

\section{Introduction}
\label{sec:introduction}

The problem of the distribution of stars in globular clusters, and more general in galactic systems, has implied one of the results of most profound interest in classical astronomy. In particular, in the pioneering works of \citet{1963MNRAS.125..127M} and \citet{1966AJ.....71...64K}, they considered the effects of collisional relaxation and tidal cutoff by studying solutions of the Fokker-Planck equation. There, it was shown that stationary solutions exist and are well described by isothermal spheres models, based on simple Maxwellian energy distributions with a constant subtracting term interpreted as an energy cutoff. An extension of this statistical analysis with thermodynamic considerations, which includes the effects of violent (collisionless) relaxation, has been studied in \citet{1967MNRAS.136..101L}, with implications to the problem of virialization in galaxies which are still of current interest \citep[see e.g.][]{2008gady.book.....B}.

Following the work of \citet{1969PhRv..187.1767R} the attention has been directed to the possible role of quantum statistics as opposed to the Boltzmannian description. Attention has correspondingly shifted from stars to elementary particles. There the case of bosons as well as fermions was considered. This also shifted the interest from the baryonic matter composing a star to a new field of interest, which has become since of great relevance, the dark matter (DM) components of galactic structures.

A first significant attempt was made by \citet{1983PhLB..122..221B} who called attention on the possible role of self-gravitating bosons for explaining galactic halos. Their result suggested as a viable DM candidate low particle masses down to $10^{-24}$~eV. This idea was further developed by a large number of authors. For a recent review of the initial work as well as the large number of intervening works see e.g. \citet{2017PhRvD..95d3541H}, and references therein.

While the works on bosons were addressing the possible smallest particle mass in nature an alternative line of research of self-gravitating fermions of masses larger than few keV, also indicated by \citet{1983PhLB..122..221B}, addresses the system of semi-degenerate self-gravitating fermions with the aim of describing galactic DM halos \citep[see e.g.][]{1990A&A...235....1G}. Further, it was considered a quantum fermionic distribution taking into account the possible presence of a cutoff in the energy as well as in the angular momentum \citep{1983A&A...119...35R,1989A&A...221....4M,1992A&A...258..223I}.
A remarkable contribution in the understanding of these issues was given in \citet{2004PhyA..332...89C}, based on the study of generalized kinetic theories accounting for collisionless relaxation processes, and leading to a class of generalized Fokker-Planck equation for fermions. It was there explicitly shown the possibility to obtain, out of general thermodynamic principles, a generalized Fermi-Dirac distribution function including an energy cutoff, extending the former Boltzmannian results by \citet{1963MNRAS.125..127M} and \citet{1966AJ.....71...64K} to quantum particles. 

More recently, it was shown that quantum particles fulfilling fermionic quantum statistics and gravitational interactions are able to successfully describe the distribution of galactic DM halos when compared with observations \citep{2013pdmg.conf30204A,2014MNRAS.442.2717D,2014JKPS...65..801A,2015ARep...59..656S,2015MNRAS.451..622R}. A similar approach to galactic halos has been developed in \citet{2015PhRvD..92l3527C} within the so-called fermionic King model, but lacking information on the fermion mass and of general relativistic effects which become important for the quantum cores approaching the critical mass for gravitational collapse. In particular, \citet{2015MNRAS.451..622R} proposed a new model (hereafter RAR model) addressing the simultaneous fulfillment of the dense quantum core to the classical halo distribution. There the RAR model was proposed as a viable possibility to establish a link between the dark central cores to DM halos within a unified approach.

%Thus, given the apparent ubiquity of massive black holes (BHs) at the center of galaxies, in \citet{2015MNRAS.451..622R} the RAR model was proposed as a viable possibility to establish: \textcolor{red}{(1) the possible explanation of the about $4\times 10^6~M_\odot$ mass concentration at the Galactic center not by a BH but by a self-gravitating system of fermionic DM particles; (2) to fulfill the properties of the flat rotation curves of galaxies due to prominent DM contribution in the halo; (3) to show the necessity of having central mass concentration in the core of galaxies in terms of a necessary condition for containing a self-consistent picture. The latter implies a link between the dark central cores to DM halos within a unified approach.}

In this paper we extend the RAR model by introducing a cutoff in the momentum distribution to account for (1) finite galaxy sizes (analogously as previously done in \citealp{1992A&A...258..223I}), and (2) to account for more realistic galaxy relaxation mechanisms as indicated above and in \citet{2004PhyA..332...89C}; providing a new family of solutions with an overall re-distribution of the bounded fermions. 

Consequently, the more stringent outer halo constraints of our novel configurations allow a higher compactness of the central cores. In fact, the possibility of a fermion core at the Galactic center as an alternative to the central BH,  first studied in \citet{2002PrPNP..48..291B} in the framework of Newtonian gravity, did not succeed in reaching the correct compactness of the quantum core since the cutoff energy parameter was not there considered.

Thus, the key questions to be answered here are the following: 
\begin{itemize}
\item 
can the gravitational potential of the new quantum core sited at the center of the DM halo be responsible for the observed dynamics of the surrounding gas and stars, without the necessity of introducing a central BH?
\item
if so, which is the allowed DM fermion mass range to account for such observational constraints?
\end{itemize}

We answer here the above questions by making a detailed analysis of the theoretical RAR DM profiles. We present in \cref{sec:rar} the details of the general relativistic equilibrium equations of this model and discuss the general features of the physical variables. Then, by using a recent and extensive observational study of the Milky Way rotation curves \citep{2013PASJ...65..118S}, and including the central S-star cluster data \citep{2009ApJ...707L.114G}, we show here (see \cref{sec:mw}) for the first time:
\begin{enumerate}
\item
That a regular and continuous distribution of keV fermions can be an alternative to the BH scenario in SgrA*, being at the same time in agreement with the Milky Way DM halo, and without spoiling the known baryonic (bulge and disk) components which dominate at intermediate scales.
\item
By constraining the DM quantum core to have the minimum compactness required by the S2 star dynamics, and by requesting the gravitational stability of the entire DM configuration, the fermion mass can be constrained to the range $mc^2 = 48$ -- $345$~keV.
\end{enumerate}

Finally in \cref{sec:conclusion} we provide a discussion of the main results of our work, and further comment on where it stands with respect to the current affairs of cosmological DM and structure formation, indicating its potentiality to solve some of the actual discrepancies within the standard $\Lambda$CDM and $\Lambda$WDM cosmologies.
\section{The Ruffini-Arg\"uelles-Rueda (RAR) model}
\label{sec:rar}

% DM model introduction
Following \citet{1992A&A...258..223I,2013IJMPD..2260008R}, we consider a system of self-gravitating massive fermions with a cutoff in the phase-space distribution under the assumption of thermodynamic equilibrium in general relativity.

A quantum phase-space function of this kind can be obtained as a (quasi) stationary solution of a generalized Fokker-Planck equation for fermions including the physics of violent relaxation and evaporation, appropriate to treat non-linear galactic DM halo structure formation \citep{2004PhyA..332...89C}. These phase-space solutions fulfill a maximization (Fermi-Dirac) entropy principle at fixed DM halo mass (bounded in radius) and temperature, consistent with the solutions given here, and further justifying the above assumed thermodynamic equilibrium approximation.

The fermionic equation of state can be written by
\begin{align}
\label{rhoepdefscutoff}
	\rho &= m\frac{2}{h^3}\int_{0}^{\epsilon_c}f_c(p)\left(1+\frac{\epsilon(p)}{mc^2}\right)d^3p\ ,\\
	P	 &= \frac{1}{3}\frac{2}{h^3}\int_{0}^{\epsilon_c}f_c(p)\,\epsilon\,\frac{1+\epsilon(p)/2mc^2}{1+\epsilon(p)/mc^2}d^3p\ ,
\end{align}
where the integration is carried out over the momentum space bounded from above by $\epsilon\leq\epsilon_c$, with $\epsilon_c$ the cutoff energy (see below); $f_c(p)$ is the phase-space distribution function differing from the standard Fermi-Dirac in the energy cutoff as
\begin{equation}
f_c(\epsilon\leq\epsilon_c) = \frac{1-e^{(\epsilon-\epsilon_c)/kT}}{e^{(\epsilon-\mu)/kT}+1}, \qquad f_c(\epsilon>\epsilon_c)=
0\, ,
\label{fcDF}
\end{equation}
where $\epsilon=\sqrt{c^2 p^2+m^2 c^4}-mc^2$ is the particle kinetic energy, $\mu$ is the chemical potential with the particle rest-energy subtracted off, $T$ is the temperature, $k$ is the Boltzmann constant, $h$ is the Planck constant, $c$ is the speed of light, and $m$ is the fermion mass. We do not include the presence of anti-fermions, i.e.~we consider temperatures $T \ll m c^2/k$. The full set of (functional) parameters of the model are defined by the temperature, degeneracy and cutoff parameters, $\beta=k T/(m c^2)$, $\theta=\mu/(k T)$ and $W=\epsilon_c/(k T)$, respectively.

We consider the system as spherically symmetric so we adopt the metric
\begin{equation}\label{eq:metric}
ds^2 = e^{\nu}c^2 dt^2 -e^{\lambda}dr^2 -r^2 d\Theta^2 -r^2\sin^2\Theta d\phi^2,
\end{equation}
where ($r$,$\Theta$,$\phi$) are the spherical coordinates, and $\nu$ and $\lambda$ depend only on the radial coordinate $r$. 

The thermodynamic equilibrium conditions are \citep{1930PhRv...35..904T,1949RvMP...21..531K}: 
\begin{eqnarray}
e^{\nu/2} T &=& {\rm constant},\label{eq:betaconstant}\\
e^{\nu/2}(\mu+m c^2) &=& {\rm constant}.
\end{eqnarray}

The cutoff condition comes from the energy conservation along a geodesic, 
\begin{equation}
e^{\nu/2}(\epsilon+m c^2) = {\rm constant},
\end{equation}
that leads to the cutoff (or \textit{escape} energy) condition 
\begin{equation}
(1+W \beta)=e^{(\nu_b-\nu)/2},
\end{equation}
where $\nu_b\equiv\nu(r_b)$ the metric function at the boundary of the configuration, i.e. $W(r_b)=\epsilon_c(r_b)=0$ \citep{1989A&A...221....4M}, and $r_b$ is the boundary radius often called \textit{tidal} radius. The above cutoff formula reduces to the known escape velocity condition $v_e^2=-2\phi$ in the classical limit $c\to\infty$ ($e^{\nu/2}\approx 1+\phi/c^2$) considered by \citet{1966AJ.....71...64K}, where $V=m\phi$ with $\phi$ the Newtonian gravitational potential, adopting the choice $V(r_b)=0$. 

The above conditions together with the Einstein equations lead to the system of equilibrium equations
\begin{align}
	\frac{d\hat M_{DM}}{d\hat r}&=4\pi\hat r^2\hat\rho, \label{eq:eqs1}\\
	\frac{d\theta}{d\hat r}&=-\frac{1-\beta_0(\theta-\theta_0)}{\beta_0}
    \frac{\hat M_{DM}+4\pi\hat P\hat r^3}{\hat r^2(1-2\hat M_{DM}/\hat r)},\label{eq:eqs2}\\
    \frac{d\nu}{d\hat r}&=\frac{2(\hat M_{DM}+4\pi\hat P\hat r^3)}{\hat r^2(1-2\hat M_{DM}/\hat r)}, \\
    \beta(\hat r)&=\beta_0 e^{\frac{\nu_0-\nu(\hat r)}{2}}, \\
    W(\hat r)&=W_0+\theta(\hat r)-\theta_0\, .\label{eq:Cutoff}
\end{align}
In the limit $W\to\infty$ (i.e. $\epsilon_c\to\infty$) these system reduce to the equations considered in the original RAR model \citep{2015MNRAS.451..622R}. We have introduced the same dimensionless quantities as in the original RAR model formulation: $\hat r=r/\chi$, $\hat M_{DM}=G M_{DM}/(c^2\chi)$, $\hat\rho=G \chi^2\rho/c^2$, $\hat P=G \chi^2 P/c^4$, where $\chi=2\pi^{3/2}(\hbar/mc)(m_p/m)$ and $m_p=\sqrt{\hbar c/G}$ the Planck mass. We note that the constants of the Tolman and Klein conditions are evaluated at the center $r=0$, indicated with a subscript `0'.

\begin{figure*}%
	\centering%
	\includegraphics[width=0.8\hsize,clip]{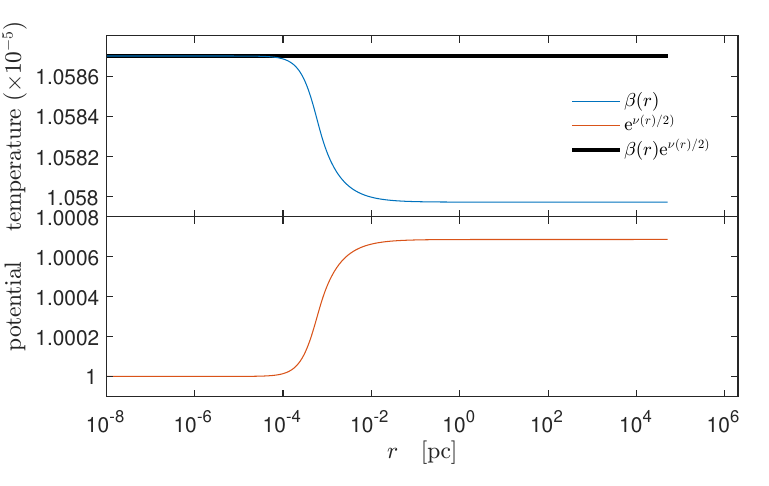}
	\caption{Upper panel: temperature parameter, $\beta$, as a function of the radial distance $r$, for a fermion mass $m c^2 = 48$~keV and for given halo boundary conditions taken from observations (see next section). We plot also the gravitational redshifted temperature $e^{\nu/2} \beta$ to check its constancy throughout the configuration as requested by the thermodynamical equilibrium condition (\protect\ref{eq:betaconstant}). Lower panel: gravitational potential, $e^{\nu/2}$.}
	\label{fig:betavsr}
\end{figure*}

\begin{figure*}%
	\centering%
	\includegraphics[width=0.8\hsize,clip]{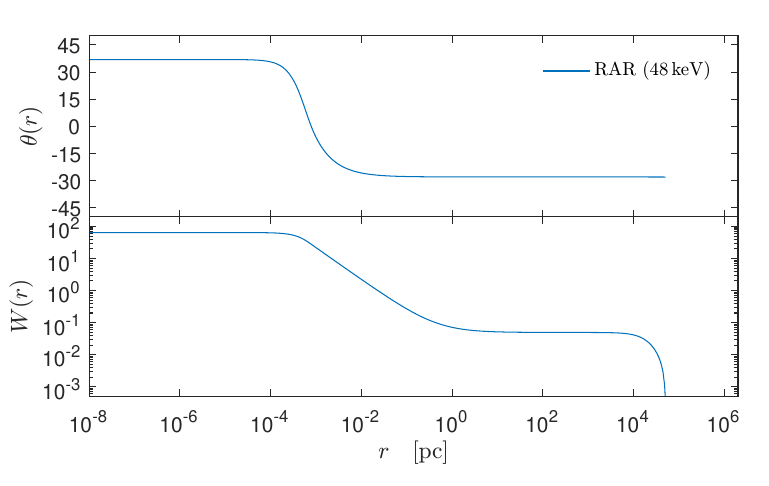}
	\caption{Upper panel: Degeneracy parameter, $\theta$, as a function of the radial distance $r$, for a fermion mass $m c^2 = 48$~keV and for given halo boundary conditions taken from observations (see next section). It can be seen the three different regimes throughout the Galaxy: the degenerate quantum core (highly positive values of $\theta$), the transition from positive to negative values where quantum corrections are still important and finally the region of highly negative values corresponding to a Boltzmannian regime. Lower panel: corresponding cutoff parameter, $W$.}
	\label{fig:thetaWvsr}
\end{figure*}

% constraints from observables
We proceed now to discuss the initial and boundary conditions for the solution of the above system of equations (\ref{eq:eqs1}--\ref{eq:Cutoff}), for given regular initial conditions at the center, $[M_{DM}(0)=0,\theta(0)=\theta_0,\beta(0)=\beta_0,\nu(0)=0,W(0)=W_0]$, for different DM particle mass $m$, to find a solution consistent with the DM halo observables of the Galaxy, which we give in next section. We show in \cref{fig:betavsr,fig:thetaWvsr} the results of the numerical integration for the radial behaviour of the free RAR model parameters for a selected fermion mass $m c^2 = 48$~keV.

In \cref{fig:betavsr} we show the temperature parameter, $\beta$, as a function of the radial distance $r$, for a fermion mass $m c^2 = 48$~keV. We plot also the gravitational redshifted temperature $e^{\nu/2} \beta$, which is constant throughout the configuration, as requested by the thermodynamical equilibrium condition in \cref{eq:betaconstant}. For completeness, we show the gravitational potential, $e^{\nu/2}$, as a function of the radial distance $r$, from which it can be checked that the temperature is higher where the gravitational potential is deeper, as given by \cref{eq:betaconstant}.

\def\ROOTPATH{figures/beta2}

\Cref{fig:thetaWvsr} shows the degeneracy parameter, $\theta$, as a function of $r$, for a fermion mass $m c^2 = 48$~keV and for given halo boundary conditions taken from observations (see next section). It can be seen the three different regimes throughout the Galaxy: the degenerate quantum core (highly positive values of $\theta$), the transition from positive to negative values where quantum corrections are still important and finally the region of highly negative values corresponding to a Boltzmannian regime.

\def\ROOTPATH{figures/thetaW}

For the same fermion mass, \cref{fig:thetaWvsr} shows the cutoff parameter as a function of $r$. From this figure it can be clearly seen the boundary of the Galaxy introduced by the cutoff (se next section for details).

With the aid of the general features of the physical RAR model parameters shown in the above figures, we can see that the DM density distribution (see ~\cref{fig:rhotot} in \cref{sec:mw}) shows in general a division of three physical regimes:
\begin{enumerate}
\item 
an inner core with radius $r_c$ of almost constant density governed by quantum degeneracy (see the region of high positive values of the degeneracy parameter in \cref{fig:thetaWvsr});
\item
an intermediate region with a sharply decreasing density distribution followed by an extended plateau, where quantum corrections are still important (see the region of transition from positive to negative values of degeneracy in \cref{fig:thetaWvsr}); and 
\item
a Boltzmannian density tail (see highly negative values of the degeneracy parameter in \cref{fig:thetaWvsr}) showing a behavior $\rho\propto r^{-n}$ with $n>2$ due to the cutoff constraint, as can be seen from  \cref{fig:thetaWvsr}. 
\end{enumerate}

Indeed, the use of the quantum (fermionic) statistical treatment is justified in all the family of inner cores analyzed here for the Galaxy since they fulfill the condition $\lambda_B\gtrsim 3 l_c$, where $l_c\sim n_c^{-1/3}$ is the interparticle mean distance within the core (with $n_c$ the core particle density) and $\lambda_B=h/(2\pi m k T_c)^{1/2}$ the thermal de-Broglie wavelength at the core (see below in \cref{sec:mw} and also \citealp{2015MNRAS.451..622R}).

The different regimes in the $\rho(r)$ profiles are also manifest in the DM rotation curve showing respectively (see \cref{fig:vrot}):
\begin{enumerate}
\item 
a linearly increasing circular velocity $v_{\rm DM}\propto r$ reaching a first maximum at the quantum core radius $r_c$; 
\item
a Keplerian $v_{\rm DM}\propto r^{-1/2}$ decreasing behavior representing the transition from quantum degeneracy to the dilute regime, continued again by a $v_{\rm DM}\propto r$ trend until reaching a second maximum at $r_h$, which we adopt as the one-halo scale length in our model; 
\item
a decreasing behavior consistent with the power law density tail $\rho\propto r^{-n}$ (with $n>2$) due to the cutoff constraint.
\end{enumerate}

The above more general \textit{dense quantum core - classical halo} distribution appears as a consequence of the inclusion of finite temperature effects together with the possibility for the fermions to acquire positive values of the degeneracy parameter in the central regions of the equilibrium configurations (regardless of an eventual cutoff in the particle energy).

The fact that the fermions are immerse in an external gravitational field automatically leads to a radial gradient of the degeneracy leading to a highly degenerate and compact core at the center followed by a sharp transition to the more diluted and extended classical Boltzmann-like tail.

It is important to stress that our results markedly differ to the ones describing galactic halos solely by a highly-degenerate configuration or solely by a classical Boltzmann-like (or diluted Fermi-Dirac) one; see e.g. \citet{2014MNRAS.442.2717D} and references therein. As we have shown both regimes exist inside the galaxy. Indeed, \citet{2017MNRAS.467.1515R} has recently shown in the specific case of dwarf galaxies that a fully degenerate self-gravitating system of fermions leads to rather compact halos in contrast with the observational values and that the problem is alleviated by the artificial addition of an isothermal density tail. Clearly, such a hybrid description is overcome by the unified core-halo description first presented in \citet{2015MNRAS.451..622R} and generalized in this work.

It is interesting to recall that a similar core-halo structure obeying a Fermi-Dirac-like distribution function was shown to arise as the most probable final outcome (from a maximization entropy principle) of collisionless relaxation mechanisms \citep{1967MNRAS.136..101L,1998MNRAS.296..569C,2004PhyA..332...89C,2015PhRvD..92l3527C}.

\section{Application to the Milky Way}
\label{sec:mw}

Based on the above morphological structure, we adopt as boundary conditions a DM halo  mass with the observed value at two different radial locations in the Galaxy: a DM halo mass $M_{\rm DM}(r=40~{\rm kpc})= 2\times 10^{11} M_\odot$, consistent with the dynamics of the outer DM halo \citep[see][and also point (i) below]{2014MNRAS.445.3788G}, and $M_{\rm DM}(r=12$~kpc)$= 5\times 10^{10} M_\odot$, as constrained in \citet{2013PASJ...65..118S}. Simultaneously, we require a quantum core of mass $M_{DM}(r=r_c)\equiv M_c= 4.2\times10^6 M_\odot$ enclosed \textit{within} a radius $r_c=r_{p(S2)}=6\times 10^{-4}$~pc, the S2 star pericenter \citep{2009ApJ...707L.114G}. This implies three boundary conditions for the three free RAR model central parameters, once the particle mass is given.

We consider here the extended high resolution rotation curve data of the Milky Way in \citet{2013PASJ...65..118S}, ranging from pc scales up to $\sim 10^2$~kpc, together with the orbital data of the eight best resolved S-cluster stars taken from \citet{2009ApJ...707L.114G}. Our analysis will thus cover in total more than nine orders of magnitude of radial extent. According to \citet{2013PASJ...65..118S}, the matter components of the Galaxy can be thus divided in 4 independent mass distributions laws, governed by different kinematics and dynamics: 
\begin{enumerate}
\item[i)]
the central ($10^{-3}\lesssim r \lesssim 2$)~pc consisting in young S-stars and molecular gas, following a Keplerian law $v\propto r^{-1/2}$, and whose dynamics is dictated by a dark and compact object of mass $M_c\approx 4\times 10^6 M_\odot$ centered in SgrA*;
\item[ii)]
an intermediate spheroidal Bulge structure ($3\lesssim r \lesssim 10^3$)~pc composed mostly of older stars, and presenting a maximum bump in the velocity curve of $v\approx 250$~km/s at $r\sim 0.4$~kpc, with inner and main mass distributions explained by the exponential spheroid model 
\begin{equation}
\rho(r)=\rho_c e^{-r/a_b};
\end{equation}
\item[iii)]
an extended flat disk ($10^3\lesssim r \lesssim10^4$)~pc including star forming regions, dust and gas, whose surface mass density is described by an exponential law 
\begin{equation}
\Sigma(R)=\Sigma_0 e^{-R/a_d}, \qquad \Sigma_0=M_d/(2\pi a_d^2),
\end{equation}
being $M_d$ the total disk mass (see \citealp{2013PASJ...65..118S} for the values of the central densities ($\rho_c$,$\Sigma_0$) and corresponding scale-lengths ($a_b,a_d$) of each baryonic (bulge+disk) model); and
\item[iv)]
a spherical halo ($10^4\lesssim r \lesssim10^5$)~pc dominated by DM and presenting a velocity peak of $v\approx 160$~km~s$^{-1}$ at about $r\sim 30$~kpc, followed by a decreasing density tail steeper than $r^{-2}$.
\end{enumerate}

Following the standard assumption in the literature that baryonic and DM do not interact each other, we have calculated the total rotation curve as 
\begin{equation}
v_{\rm rot}=\sqrt{r \frac{d\Phi_T}{dr}}=\sqrt{v^2_b(r)+v^2_d(r)+v^2_{\rm DM}(r)},
\end{equation}
where
\begin{equation}
\Phi_T=\Phi_b+\Phi_d+\Phi_{\rm DM},
\end{equation}
is the total gravitational potential generated by the sum of each component, and $v^2_b(r)$, $v^2_d(r)$ the baryonic squared circular velocities. We calculated the total (inner $+$ main) bulge circular velocity using the same mass model parameters as in \citet{2013PASJ...65..118S}. For the disk, we have performed the calculations with mass models parameters ($M_d,r_d$) slightly changed with respect to those given in \citet{2013PASJ...65..118S}, where the NFW DM profile was assumed. Such slight changes imply a shift in the disk velocity of up to $15\%$ respect to the disk velocity model used in \citet{2013PASJ...65..118S}, corresponding with a $\sim 20$~km/s maximum deviance. We do this change to improve the fit of the observational data when adopting our DM profile. Finally, $v^2_{\rm DM}(r)$ is the DM contribution computed numerically from the $M_{DM}(r)$ solution of (\ref{eq:eqs1}--\ref{eq:Cutoff}), through the general relativistic formula for the velocity 
\begin{equation}
v_{DM}^2(r)=\frac{G M_{DM}(r)}{r-2 G M_{DM}(r)/c^2},
\end{equation}
according to the equations of motion of a test-particle in the spacetime metric (\ref{eq:metric}).

We discuss in next the results of the numerical integration accounting for the full boundary condition problem in the case of the Milky Way within our model. The key result is that there is a continuous underlying DM distribution covering the whole observed Galactic extent, which not only governs the dynamics of the outer halo ($r\gtrsim 10$~kpc), but also the central regions of the Galaxy ($r \lesssim 1$~pc), while the intermediate region ($1$~pc~$\lesssim r\lesssim 10$~kpc) is dominated by the baryonic components (bulge+disk).

\begin{figure*}%
	\centering%
	\includegraphics[width=0.7\hsize,clip]{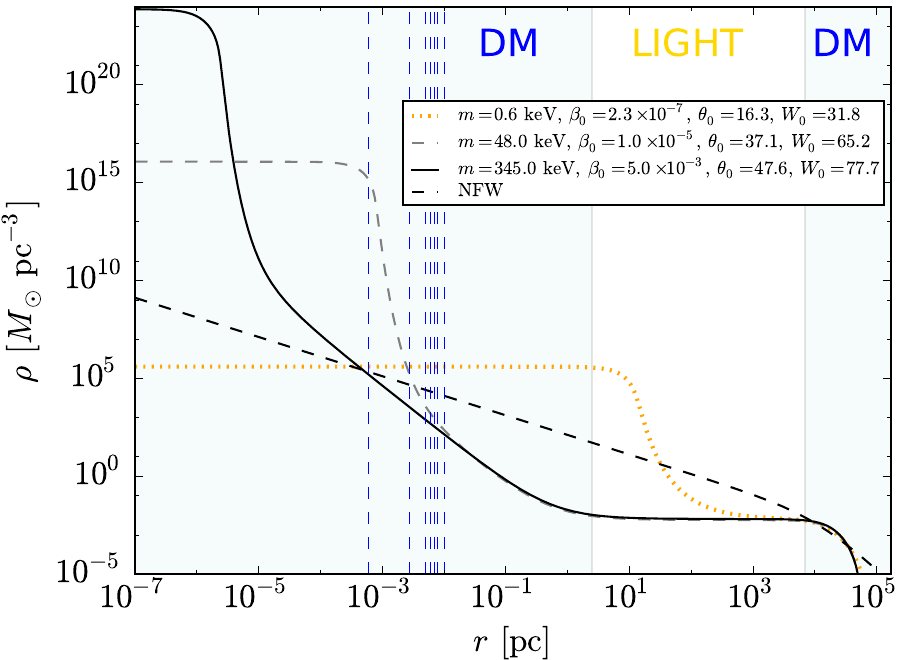}
	\caption{(Color online) Theoretical density profiles from $10^{-7}$~pc all the way to $10^5$~pc, for three representative fermion masses in the $mc^2\sim$keV region: 0.6~keV (dotted yellow curve), 48~keV (long-dashed gray curve) and 345~keV (solid black curve). The dashed blue lines indicate the position of the S-cluster stars \protect\citep{2009ApJ...707L.114G}. We show for the sake of comparison the NFW density profile as obtained in \citet{2013PASJ...65..118S} (dashed black curve). See the text for additional details.}
	\label{fig:rhotot}
\end{figure*}

\Cref{fig:rhotot} shows the RAR density profiles from $10^{-7}$~pc all the way to $10^5$~pc, for three representative values of the fermion mass: 0.6~keV$/c^2$ (dotted yellow curve), 48~keV$/c^2$ (long-dashed gray curve) and 345~keV$/c^2$ (solid black curve). The dashed blue vertical lines indicate the position of the best resolved stars of the S-cluster \citep{2009ApJ...707L.114G}. We show for the sake of comparison the NFW density profile as implemented in \citet{2013PASJ...65..118S} (dashed black curve). Generalized NFW profiles can be also used for a comparison with more recent results, considering these phenomenological profiles have been recently implemented and shown to increase the degree of compatibility with the Milky Way rotation curve \citep[see e.g.][]{2015JCAP...12..001P}.

\def\ROOTPATH{figures/rhotot}

The Milky Way outermost DM halo behavior is subjected to the cutoff conditions: $W(r_b)\approx 0$ when $\rho(r_b)= 10^{-5} M_\odot/pc^3$, at the boundary radius $r_b= 50$~kpc (see \cref{fig:rhotot}); with $\rho(r_b)$ the Local Group density as constrained in \citet{2012PASJ...64...75S}. We note that the exact $W(r)=0$ cutoff condition, is fulfilled in the limiting case $\rho(r)=0$ achieved for $r\gtrsim r_b$. The limiting behavior of such a DM density profile is also consistent with a DM halo mass of $M_{\rm DM}(r=40$~kpc)$= 2\times 10^{11} M_\odot$ as required above, further implying a total Galactic mass (dark $+$ baryonic) at $r_b$ of $M_T(r_b)\approx 3\times 10^{11} M_\odot$, of which $80\%$ is DM according to our model (\textbf{i.e.} $M_{DM}(r_b)=2.4\times10^{11} M_\odot$). It is clear that such a DM mass distribution must be also in agreement with the dynamical constraints set by the Galactic satellite dwarf observations, e.g.~the Sagittarius (Sgr) dwarf satellite. Indeed, such observational constraints have been recently considered in \citet{2014MNRAS.445.3788G}, who showed that their fulfillment requires a total mass of the Galaxy (at $\sim 80\%$ confidence level) $M_{T}(r\gtrsim 50$~kpc)$\approx 3\times 10^{11} M_\odot$, in agreement with our results here. \footnote{Constraints on the total (virial) Galaxy mass from the Sgr dwarf stream \citep{2014MNRAS.437..116B} may imply even larger values of $M_{T}(r=100$~kpc)$\approx 4\times 10^{11} M_\odot$ \citep{2014MNRAS.445.3788G}. Nevertheless, this stream motion of tidally disrupted stars is likely related with merging processes that date back to the DM halo formation of the Galaxy \citep{1995MNRAS.275..429L}, while our modeling does not include mergers, nor dynamical DM accretion from environment, which may likely increase the Galaxy mass during its whole evolution.}

\begin{figure*}%
	\centering%
	\includegraphics[width=0.7\hsize,clip]{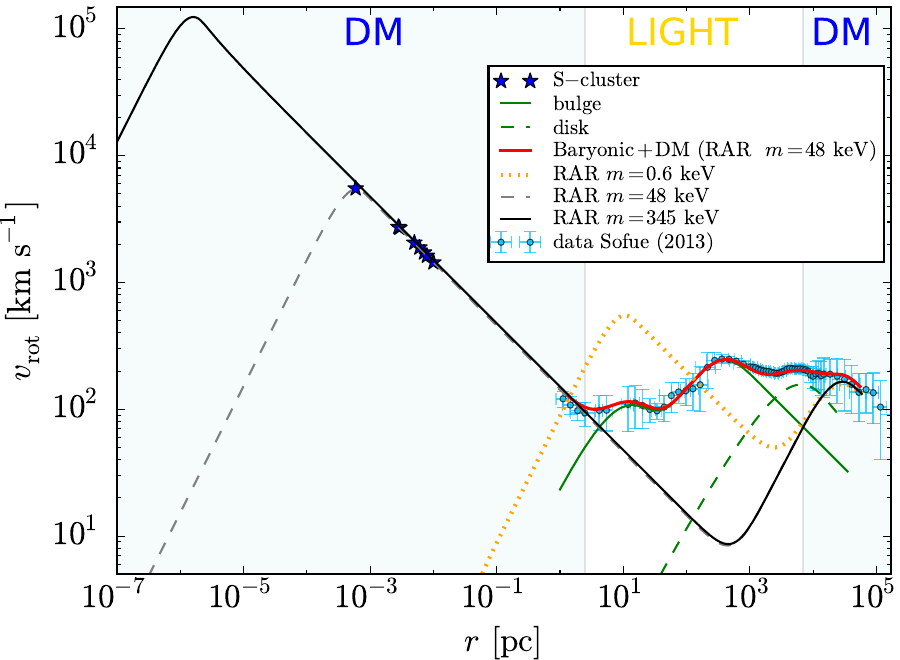}
	\caption{(Color online) Theoretical RAR rotation curves from $10^{-7}$~pc all the way to $10^5$~pc, for three representative fermion masses in the $mc^2\sim$keV region: 0.6~keV (dotted yellow curve), 48~keV (long-dashed gray curve) and 345~keV (solid black curve). These RAR solutions are in agreement with all the Milky Way observables from $\sim 10^{-3}$~pc to $\sim 10^{5}$~pc. For the case of $mc^2=48$~keV, we include the total rotation curve (red thick curve) including the total baryonic (bulge + disk) component. The star symbols represent the eight best resolved S-cluster stars \protect\citep{2009ApJ...707L.114G}. See the text for additional details.}
	\label{fig:vrot}
\end{figure*}

\Cref{fig:vrot} shows the RAR rotation curves in the same radial extent and for the same three representative values of the fermion mass in \cref{fig:rhotot}. We show for the above three fermion masses the DM contribution to the total rotation curve, and, for the case of $mc^2=48$~keV, the total rotation curve (red thick curve) including the total baryonic (bulge + disk) component. From these three examples shown, it can be directly seen that only the RAR solutions with particle masses in the range $mc^2 = 48$ -- $345$~keV are in agreement with the Milky Way observables in the region $r\sim 10^{-3}$ -- $10^{5}$~pc, where data are available. This is, the ones which are able to provide an alternative to the BH scenario in SgrA*. The blue star symbols represent the best resolved stars of the S-cluster \citep{2009ApJ...707L.114G} and their position in the plot has been evaluated as the \textit{effective} circular velocity at pericenter, i.e. without considering the ellipticity of the orbits. 

\def\ROOTPATH{figures/vrot}

\begin{figure*}%
	\centering%
	\includegraphics[width=0.7\hsize,clip]{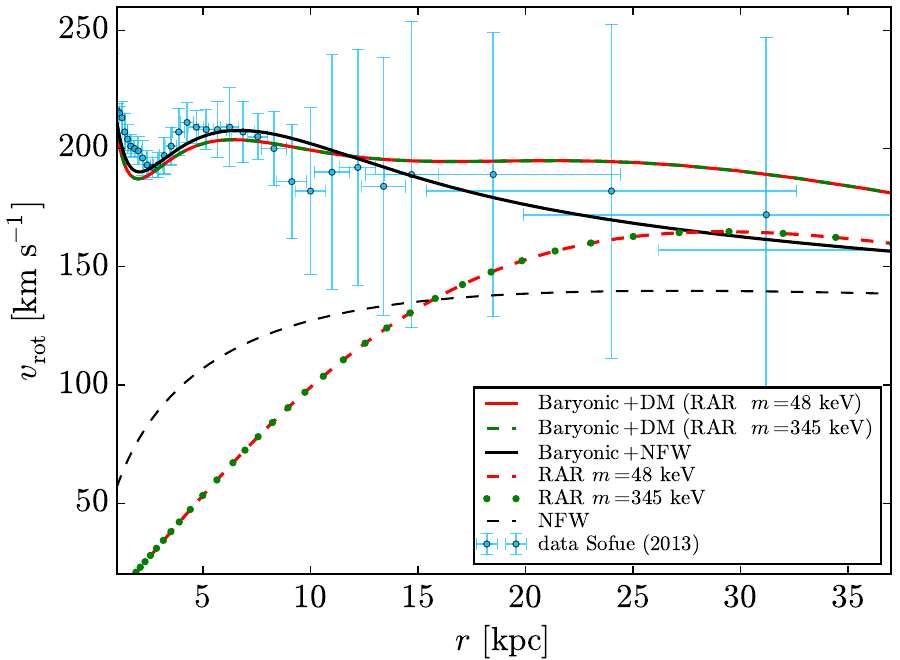}
	\caption{(Color online) Comparison of the total (DM plus baryonic) rotation curves as given by the RAR model (for a fermion mass $m c^2=48$~keV and 345~keV) and the phenomenological NFW model (from \protect\citet{2013PASJ...65..118S}, but see also \protect\citet{2015JCAP...12..001P} for the possible implementation of generalized NFW models) in the region $r = 1$ -- $35$~kpc. Within the region $1$ -- several~kpc, the contribution of the baryonic components dominates respect to the DM, while at distances $r\gtrsim 10$~kpc there is, instead, an increasing dominance of the DM component (see text for further details).}
	\label{fig:RARNFWzoomvelocity}
\end{figure*}

We further display in \cref{fig:RARNFWzoomvelocity} a zoom of \cref{fig:vrot} (the most relevant curves) in the $1-35$~kpc region and in linear scale. This allow us to better appreciate the difference between the diverse DM models in the radial window where the Rotation Curve is most relevant. Thus, \cref{fig:RARNFWzoomvelocity} compares the [baryon $+$ DM] Rotation Curve fits against data, for the following DM models: (i) NFW as implemented in \citet{2013PASJ...65..118S} (continuous black); (ii) RAR models for $mc^2=48$~keV (solid red) and $mc^2=345$~keV (dashed green). We further plot in the same figure the DM-only contribution to the corresponding total Rotation Curve for each DM model.

As already evidenced in \cref{fig:vrot}, and due to the specific MW halo boundary conditions imposed to all the RAR solutions worked out in this paper (see above), there is an almost perfect match (within 1$\%$) in all RAR curves from inner-halo scales and beyond (i.e within the increasing trend $v_{DM}\propto r$ and up to $r_b$). In particular, the coinciding behaviour between the two ($mc^2=48, 345$~keV) different DM RAR profiles in the zoomed-in region, together with the addition of the unchanged baryonic contributions, explains the coinciding total (baryon $+$ RAR-DM) Rotation Curve fits (solid red and dashed green).

Instead, there is a visible and appreciable difference when comparing between the (intrinsically) distinct RAR and NFW (DM-only) models, mainly in the region of $1$ -- $10$~kpc. Nevertheless, and regardless of this discrepancy, both profiles provide a comparably good fit to the total Rotation Curve, given the baryonic physics (bulge $+$ disk) is the one which dominate at those scales. While the slight exceeding trend in total $v_{rot}$ of [baryon $+$ NFW-DM] (black curve) respect to [baryon $+$ RAR-DM] (e.g. red curve)\footnote{
	Slight changes in the disk parameters where applied within RAR model-fitting respect to those given in \citet{2013PASJ...65..118S} where NFW was applied, as explicited above in this section.
} is explained by the cuspiness of the NFW (inner-halo) density profile respect to the cored RAR profile at those scales. At outer radii, $r > 10$~kpc, DM dominates over baryons, and now the exceeding trend in $v_{rot}$ of [baryon $+$ RAR-DM] over [baryon $+$ NFW-DM] is directly explained by the difference in the power-law behaviours of NFW ($\rho\propto r^{-3}$), respect to RAR ($\rho\propto r^{-n}$), with $n>2$.

Finally, it is important to stress that we have just applied a plain fitting procedure to the data (i.e. two fixed halo boundary conditions for DM RAR equations, with given simple baryonic models taken from literature), and that the only outcome of applying any more sophisticated statistical method (such as a Monte Carlo Markov Chain, or similar) is of a further improvement of our fittings.

\def\ROOTPATH{figures/vrotzoom}

% \Cref{fig:RARNFWzoomvelocity} shows the total (DM+baryonic) rotation curve as produced by the RAR ($m c^2=48$~keV) as well as the (NFW+baryonic) velocity curve, in the region $r = 2$ -- $40$~kpc, in linear scale. It can be seen in the region $1$ -- $10$~kpc, the contributions of the baryonic and DM components are comparable (in both models) while, at distances $r\gtrsim 10$~kpc, there is an increasing dominance of the DM component. 

We summarize as follows the more general results worked out in this paper, including the examples shown in the figures below:
\begin{enumerate}
\item
The fermion mass range $mc^2 \lesssim 10$~keV is firmly ruled out by the present analysis because the corresponding rotation curve starts to exceed the total velocity observed in the baryonic (bulge) dominated region $r \approx 2$ -- $100$~pc (including upper bound in error bars; see for example the highly exceeding case of $mc^2 \sim 1$~keV (dotted yellow curve) in \cref{fig:vrot}). For these particle masses the DM distribution produces a large overshoot over the observed inner-rotation curve. This implies that our lower limit to the fermion mass will hold also for different and more accurate inner-baryonic models \citep[e.g.][]{2017MNRAS.465.1621P} which, in any case, change the total inner-rotation curve only by a small percentage with respect to the one we have used in this work. In addition, for these relatively low particle masses below $10$~keV, and due to the overshooting in the inner-bulge velocity region, it is clear that these solutions only fulfill with the chosen halo boundary conditions and do not provide an alternative to the central BH in SgrA*.
\item
All RAR model solutions for a fermion mass in the range $mc^2 \approx 10$ -- $345$~keV give a nearly equal total rotation curve (as the red thick curve shown in \cref{fig:vrot} for the case of $mc^2=48$~keV) in the aforementioned baryonic dominated region above $\sim 2$~pc, since the DM potential there produces a negligible contribution with respect to the baryonic one (see also point 9 below).
\item
In the intermediate fermion mass range $mc^2 = 10$ -- $48$~keV, the theoretical rotation curve is not in conflict with any of the observed data and DM inferences in \citet{2013PASJ...65..118S}, but the compactness of the quantum core is not enough to be an alternative to the central BH scenario in SgrA*.
\item
For fermion masses $mc^2 = 48$ -- $345$~keV, the RAR solutions with corresponding initial parameters ($\beta_0$, $\theta_0$, $W_0$) explain the Galactic DM halo and, provide at the same time, an alternative for the central BH scenario. The mass lower bound in $m$ is imposed by the dynamics of the stellar S-cluster. Namely, the quantum core radius of the solutions for $mc^2 \geq 48$~keV are always smaller or equal than the radius of the S-2 star pericenter, i.e. $r_c \leq r_{p(S2)} = 6\times 10^{-4}$~pc~$\approx 1.5\times 10^3 r_{\rm Sch}$.
\item
There is a mass upper bound of $mc^2 = 345$~keV that corresponds to the last stable configuration before reaching the critical mass for gravitational collapse ($M_c^{cr}\propto m_{\rm Planck}^3/m^2$), and calculated following the turning point criterion for core-collapse in \citealp{2014IJMPD..2342020A}. The core radius of the critical configuration is $r_c\approx 4 \,r_{\rm Sch}$, with $r_{\rm Sch}$ the Schwarzschild radius of a $4.2\times10^6 M_\odot$ BH \citep[see also][]{2014JKPS...65..809A}. In particular the following set of initial conditions
$$[\beta_0=1.03\times10^{-5},\theta_0=37.14,W_0=65.252],$$
$$[\beta^{cr}_0=5.04\times10^{-3},\theta^{cr}_0=47.59,W^{cr}_0=77.706],$$ were obtained for the lower and upper DM particle mass bounds respectively (the upper index `cr' stands for the critical configuration).
\item
For a relatively low compact core as the one calculated for $mc^2=10.4~{\rm keV}$, we have $\lambda_B=3.1\,l_c$; while in the most compact one (black solid curve in \cref{fig:rhotot} for $mc^2=345$~keV) we have $\lambda_B=4.0\, l_c$. This justifies our use of the quantum fermion statistical treatment.
\item
The DM contribution to the Galactic halo rotation curves becomes necessary above $\sim 7$~kpc (see \cref{fig:vrot,fig:RARNFWzoomvelocity}). This is in agreement with the DM model-independent observational analysis by \citet{2015NatPh..11..245I}.
\item
Interestingly, in the mass range $mc^2 = 10$ -- $345$~keV the RAR DM distribution predicts Keplerian rotation curves at $r\lesssim 2$~pc (see \cref{fig:vrot}). This feature is in agreement with the apparent Keplerian trend observed in the innermost gas data points in \citet{2013PASJ...65..118S}, with the caveat that the tracers in the region $1$ -- $100$~pc might not follow the gravitational potential due to the more complex bar pattern speed present in the Milky Way central region (see also the discussion section in \citet{2013PASJ...65..118S}).
\item
In the above mass range the full rotation curve $v_{\rm rot}=\sqrt{r (d\Phi_T/dr)}$ (see solid red line in \cref{fig:vrot}) is in good agreement with observations within the observational errors. In addition, the minimum in the DM rotation curve \textit{coincides} with the absolute maximum of $v_{\rm rot}$ (i.e. the bulge peak) attained at $r\approx 0.4$~kpc. This peculiar fact might provide a clue for a deeper understanding of the complex ensemble history of the baryonic stellar bulge on top of the previously formed DM structure. This inference, as the previous one, relies on the assumption that the tracers in such a Galaxy region follow the gravitational potential, and therefore such a link must be taken with caution considering the lack of axisimmetry of the potential at those bulge scales (see however \citet{2013PASJ...65..118S} where such effects were further discussed).
\end{enumerate}

\section{Concluding remarks}
\label{sec:conclusion}

It is now clear from our results that gravitationally bounded systems based on fermionic phase-space distributions including a particle energy cutoff (or escape velocity effects) and central degeneracy, can explain the DM content in the Galaxy while providing a natural alternative for the central BH scenario in SgrA*. 

Specifically, we have shown that:
\begin{itemize}
\item
fermion masses $mc^2 \lesssim 10$~keV are ruled out by the Milky Way rotation curve since the contribution of the DM to the rotation curve highly exceeds the observed values (see \cref{fig:vrot} in \cref{sec:mw});
\item
for $mc^2 = 48$ -- $345$~keV, the DM halo distribution is in agreement with standard data of the Milky Way rotation curve tracers, and harbors a dense quantum core of $4\times10^6 M_\odot$ within the S2-star pericenter;
\item
for $mc^2 = 10$ -- $48$~keV the DM distribution is also in agreement with the Galaxy rotation curve data but the compactness of the DM quantum core is not enough to explain the S-cluster dynamics and so to be an alternative to the central BH scenario.
\end{itemize}

Our general results on the DM distribution in galaxies can be considered as complementary to those based on standard cosmological simulations. However, the latter being based on N-body purely Newtonian simulations, contrast with our semi-analytical four-parametric approach which has the chance to include more rich physical ingredients, such as quantum statistics (arising from specific phase-space relaxation mechanisms), thermodynamics, and gravity. 
%We have shown that our density profiles successfully agree with a large variety of galactic observables and universal laws.

An important aspect of the particle mass range of few $10$ -- $100$~keV obtained here from galactic observables, is that it produces basically the same behavior in the power spectrum (down to Mpc scales) from that of standard $\Lambda$CDM cosmologies, thus providing the expected large-scale structure \citep[see][for details]{2009ARNPS..59..191B}. In addition, it is not `too warm' (i.e. our masses are larger than $mc^2\sim 1$ -- $3$~keV) to enter in tension with current Ly$\alpha$ forest constraints \citep{2009PhRvL.102t1304B,2013PhRvD..88d3502V} and the number of Milky Way satellites \citep{2008ApJ...688..277T}, as in standard $\Lambda$WDM cosmologies.

Our model naturally provides cored inner DM halos (our fermionic phase-space distributions imply an extended plateau in the DM density profile on halo scales in a way that they resemble Burkert or cored Einasto profiles \citep{2015MNRAS.451..622R}), without developing undesired cuspy density trends on such scales as the ones found in N-body simulations \citep{1997ApJ...490..493N}. Such a marked difference between both kind of density profiles, the cored (RAR) and cuspy (NFW), arises due to the physics involved in the two different approaches, which in turn may provide an important insight to one of the main open problems in standard $\Lambda$CDM cosmology, i.e.~the so-called core-cusp problem \citep{2010AdAst2010E...5D}. 

Interestingly, in \citet{2011ApJ...742...20W} it has been provided a clear observational evidence for the cored nature (DM model independent) of the DM density profiles in dwarf spheroidal galaxies such as Sculptor. We show in an accompanying article (Arg\"uelles, et al. 2018; to be submitted) that these DM halo features, i.e. the constant inner halo density and halo scale-radius, are in agreement with the RAR model results applied to dwarf galaxies.

Moreover, in the accompanying article (Argüelles, et al. 2018; to be submitted), we show that a key point of the present RAR model is the ability to predict entire DM halo configurations which fulfills the observed universal properties of galaxies, such as the \textit{central BH mass - total halo mass} ($\MBH$-$\Mtot$) and the surface DM halo density relation ($\Sigma_{0D}\approx$~constant), for a unique DM fermionic mass. At the same time, it provides, on astrophysical basis, possible clues on the formation of supermassive BHs in galactic nuclei.

All the above offer significant support for our keV-scale fermions as DM, which may well co-exist harmonically with other DM species in the universe. These aspects will have to interplay with the physics of elementary particles regarding the nature of these fermions: Majorana neutrinos, supersymmetric particles, sterile neutrinos, etc.; as well as with the possible detection through decaying processes involving weak interactions. Indeed, DM fermion masses within the relatively narrow window obtained here, $mc^2= 48$ -- $345$~keV, have also arisen within different microscopic models based on extensions of the standard model, and consistent with all cosmological, large scale structure, and X-ray constraints, as the ones considered in \citet{2009PhRvL.102t1304B,2015PhRvD..92j3509P}.

\section*{Acknowledgments}

%We thank the referee for her/his very constructive and clear suggestions.
%
%C.R.A acknowledges support by the International centre for Relativistic Astrophysics Network (ICRANet) and CONICET-Argentina.
%
%J.A.R acknowledges support from the International Cooperation Program CAPES-ICRANet financed by CAPES-Brazilian Federal Agency for Support and Evaluation of Graduate Education within the Ministry of Education of Brazil.
%
We thank the referee for the very constructive and clear suggestions. A.K. is supported by the Erasmus Mundus Joint Doctorate Program by Grants Number 2014--0707 from the agency EACEA of the European Commission.

\bibliographystyle{elsarticle-num-names}
\bibliography{biblio}

\end{document}